\documentclass{aa}

\usepackage{graphicx}
\usepackage{txfonts}
\usepackage{paralist}
 
\begin{document}

\title {Chemical evolution of the Galactic bulge: different stellar populations and possible gradients}

\author {V. Grieco\inst{1} \thanks {email to:
    grieco@oats.inaf.it}\and F. Matteucci\inst{1,2}\and A. Pipino\inst{3}\and
G. Cescutti\inst{4}
} 
 \institute{Dipartimento di Fisica, Sezione di Astronomia, 
Universit\`a di Trieste, via
  G.B. Tiepolo 11, I-34131, Trieste, Italy \and I.N.A.F. Osservatorio
  Astronomico di Trieste, via G.B. Tiepolo 11, I-34131, Trieste,
  Italy \and Institut fur Astronomie, ETH Zurich, Wolfgang-Pauli-Str.27, 8093 Zurich, CH \and Leibniz-Institut f\"ur Astrophysik Potsdam (AIP), An der Sternwarte 16, 14482, Potsdam, Germany
}

\date{Received xxxx / Accepted xxxx}

\abstract {The recent although contoversial discovery of two main stellar populations in the Galactic bulge, one metal poor, with a spheroid  kinematics and the 
other metal rich, with a bar-like kinematics, suggests to revise the classical model for bulge formation.}
{We aim at computing  the chemical evolution of the Galactic bulge to explain the existence of the two main stellar populations. 
We also plan to explore the possible existence of spatial abundance gradients inside the bulge.} 
{We adopt a chemical evolution model which follows the evolution of several chemical species (from H to Ba). 
We assume that the metal poor population formed first and on a short timescale  while the metal rich population formed later and out 
of the enriched gas. We predict the stellar distribution functions for Fe and Mg, 
the  mean $<[Fe/H]>$ and $<[Mg/H]>$ and the [Mg/Fe] vs. [Fe/H] relations in the two stellar populations. We also consider the case in which the metal poor 
population could be the result of sub-populations formed with different chemical enrichment rates. }
{Our results , when compared with observations, indicate that the old more metal poor stellar population formed very fast (on a timescale of 0.1-0.3 Gyr) by means of an 
intense burst of star formation and  an initial mass function flatter than in the solar vicinity. 
The metal rich population formed on a longer timescale (3 Gyr). We predict differences in the mean abundances of the two populations (-0.52 dex for $<[Fe/H]>$)
which can be interpreted as a metallicity gradients. We also 
predict possible gradients for Fe, O, Mg, Si, S and Ba between sub-populations inside the metal poor population itself (e.g. -0.145 dex  for  $<[Fe/H]>$). 
Finally, by means of a chemo-dynamical model following a dissipational collapse, we predict a gradient inside 500 pc from the Galactic center of $-0.26 dex kpc^{-1}$ in Fe.}
{A stellar population forming by means of a classical gravitational gas collapse is probably mixed with a younger stellar population created perhaps by the bar evolution. 
The differences among their mean abundances can be interpreted as gradients. On the basis of both chemical and chemo-dynamical models, 
we also conclude that it is possible that the metal poor population contains abundance gradients itself, and therefore different stellar populations.}


\titlerunning{   }
\authorrunning{  }
\maketitle
 
\section{Introduction}
Recent data concerning the Galactic bulge are suggesting a rather complex picture for its formation. 
Recently, Babusiaux et al. (2010), Gonzalez et al. (2011), Bensby et al. (2011), 
 Hill et al. (2011)and  Robin et al. (2012) have shown that the stellar populations and chemical evolution of the Galactic bulge 
are not as simple as it could have appeared up to now. 
In particular, Bensby et al. (2011) observed microlensed dwarfs and subgiant bulge stars and concluded that their distribution is bimodal 
with one peak at [Fe/H]=-0.6 dex and one peak at [Fe/H]=0.3 dex. Hill et al. (2011) by studying bulge red clump stars also found two 
distinct stellar populations in the bulge, one with a peak at [Fe/H]$\sim$ -0.45 dex and another with a peak at [Fe/H]$\sim$ +0.3 dex. 
The interpretation of these two populations is that the metal poor (MP) one probably reflects the classical bulge component, the old 
spheroid population formed in a short timescale, as witnessed by the high [Mg/Fe]$\sim$ +0.3 dex ratio and the kinematics corresponding 
to an old spheroid (Babusiaux et al. 2010), whereas the metal rich (MR) population seems  to possess a bar kinematics and it could have originated by a pre-enriched gas coming 
either from the residual gas from the formation of the metal poor component or from the metal rich inner disk. These stars could have formed 
on a longer timescale than the metal poor component, as witnessed by their  almost solar [Mg/Fe] ratios.
In fact, as it is well known, in a regime of a very fast star formation rate, most of the stars form with high [Mg/Fe] ratios, due to the 
predominant pollution by core-collapse supernovae (SNe). On the other hand, in a regime of slow star formation, the [Mg/Fe] ratios tend to
 be lower, due to the pollution by Type Ia SNe intervening later than core-collapse SNe in the chemical enrichment process.

What emerges from the recent data is therefore that the Galactic bulge could have both the characteristics of a classical bulge and a pseudo-bulge. On one side, the existence of a bar is now proven by several studies (e.g. McWilliam \& Zoccali, 2010; Saito et al. 2011) suggesting that the bulge has an X-shaped structure, which can indicate the existence of a bar. The extensive survey by 
Ness \& Freeman (2012) has shown that the Milky Way bulge is indeed a bar. Results from BRAVA survey (e.g. Rich \& al. 2007) did not find evidence for a different population from the bar one, whereas Shen \& al. (2010) and Kunder \& al. (2011) suggested that the classical bulge component exists but it is $< 8\%$ of the mass of the disk. 
On the other hand, color-magnitude diagram analyses of bulge stars (e.g. Zoccali et al. 2003; Kuijken \& Rich 2002; Clarkson et al. 2008; 2011) have  indicated that the bulge is old and that there is a little age spread among stars. This fact, coupled with the high [Mg/Fe] ratios, argues in favor of a fast bulge formation . Therefore, the situation seems to be quite complex and even contradictory. 

Various theories for the bulge formation were put forward in the past years. Wyse \& Gilmore (1992) first summarized the various possibilities and in the following years many studied appeared on the subject. The main proposed scenarios are as follows:

\begin{itemize}
\item a) {\bf Accretion of stellar satellites}. This idea was later developed in models were the bulge formed by accretion of extant stellar systems which hierarchically merged and eventually settled in the center of the Galaxy (Noguchi, 1999; Aguerri et al. 2001; Bournaud et al. 1999);

\item b) {\bf In situ star formation from primordial or slightly enriched gas.}
The bulge was formed by a fast gravitational collapse (Larson, 1976) or slow accumulation of gas at the center of the Galaxy and subsequent evolution with either fast or a slow star formation; the accreting gas could have been primordial or metal enriched by the halo, thick-disk or thin-disk. In the past years, the comparison between model predictions and the observed metallicity distribution function (MDF) of the Galactic bulge suggested that this component of the Milky Way had evolved very fast and with a flatter initial 
mass function (IMF) than in the solar vicinity (e.g. Matteucci \& Brocato, 1990;
Matteucci et al. 1999; Ferreras et al. 2003; 
Ballero et al. 2007; Cescutti \& Matteucci, 2011).

\item c)  {\bf Secular evolution}. The bulge formed as a result of secular evolution of the disk through a bar forming a pseudo-bulge (e.g. Combes et al. 1990; Norman et al. 1996; Kormendy \& Kennicutt (2004); Athanassoula, 2005). After the formation of the bar, the bulge heats in the vertical direction giving rise to the typical boxy/peanut configuration. A more recent model belonging to this category assumes that the bulge is formed though bas instability from a disk composed by thin and thick disk components (Bekki \& Tsujimoto, 2011). However, these models were not tested on the observed chemical abundances.

\item d)  {\bf Mixed scenario : secular and spheroidal components together.}
Samland \& Gerhard (2003) had predicted, by means of a dynamical model, the existence of two bulge populations: one formed in an early collapse
 and the other formed late in the bar.
Although a two-step formation of the bulge is not a new idea (see Wyse \& Gilmore 1992), recently, Tsujimoto \& Bekki (2012) tried to model the two main stellar populations found in the bulge and suggested that the metal poor 
component formed on a timescale of 1 Gyr with a flat IMF (x=1.05), whereas the other component, the metal rich one, formed from pre-enriched gas on a timescale of 4 Gyr.

\end{itemize}

In this paper we aim at computing the chemical evolution of the Galactic bulge by means of a very detailed chemical evolution model following the evolution of 36 chemical species, and see whether we are able to reproduce the two observed main stellar components (Hill et al. 2011) and their abundances under reasonable assumptions. In particular, we aim at  testing whether the MP population can be explained by a less flat IMF than suggested by Ballero et al. (2007) and Cescutti \& Matteucci (2011), on the basis of the previous observed stellar metallicity distribution functions available up to now. Then we will study the differences among the mean abundances of several chemical elements in the two populations. In principle, these differences can be interpreted as abundance gradients, although the two populations are likely to be spatially mixed.

Moreover, we intend to compute possible abundance gradients inside the MP population and see whether they are compatible with the data. The observational situation is, in fact, not yet clear.
Rich et al. (2012) did not find any  vertical gradient from the Galactic center to the Baade's window, inside the innermost 600 pc. If true, this suggests that the Baade's window stellar population formed indeed very fast so that no gradient could be created. In fact, an abundance gradient is naturally created during a dissipative collapse (Larson 1976). On the other hand, Zoccali et al. (2008;2009) found different mean Fe abundances by analysing different fields in the bulge: in the Baade's window field they found $<[Fe/H]> =+0.03$ dex at  b=$-4\,^{\circ}$,  while in a field at higher latitude (b=$-12\,^{\circ}$) they found $<[Fe/H]> =-0.40$ dex. This difference can clearly be interpreted as a  vertical gradient along the bulge minor axis.

Pipino et al. (2008) run 1D chemo-dynamical models for a  Milky Way-like bulge ($M_* \sim 2 \cdot 10^{10}M_{\odot}$ and $R_{eff}=$1 kpc) and found that during the gravitational collapse giving rise to the bulge, an abundance gradient in the stars is indeed created. In particular, they suggested that inside 1 kpc from the Galactic center we should expect a gradient in the global stellar metallicity of d[Z/H]/dR=-0.22dex $kpc^{-1}$. 
Here, we will rerun this model and give predictions for the gradients of O and Fe. Moreover, in a simple way we will explore if inside the MP bulge 
component we can identify at least two sub-populations, formed at different 
rates and showing different average abundances.

Finally, it is worth noting that we will  concentrate on explaining the observed metallicity distribution and chemical abundances of Galactic bulge stars and that we cannot say much about secular evolution and bar formation, since our model does not take into account stellar dynamics. Galactic chemical evolution can only put constraints on the timescales of formation of the different stellar populations, but it cannot predict how the bulge actually formed. For this reason, in this paper we are concerned only with chemical abundances.

The paper is organized as follows: in Section 2 we describe the observational data , in Section 3 the chemical evolution model and  in Section 4 the results are presented and compared with observations. Finally, in Section 5, we present a discussion on bulge formation and some conclusions are drawn.

\section{The observational data}

In this paper we compare our results with the recent data from 
Hill et al. (2011).
In that paper they presented measures of [Fe/H] for 219 bulge red clump stars from R=20000 resolution spectra obtained with FLAMES/GIRAFFE at the VLT. For a subsample of 162 stars they measured also [Mg/H].The stars are all in a Baade's window. They interpreted the iron distribution in bulge stars as bimodal, indicating two different stellar components of equal size: a metal poor component centered  around [Fe/H] =-0.30 dex and [Mg/H]=-0.06 dex with a large dispersion and a metal-rich narrow component centered around [Fe/H]=+0.32dex  and [Mg/H]= +0.35 dex. Therefore the metal poor component shows high average [Mg/Fe] $\sim$ 0.3 dex, whereas the metal rich one  shows [Mg/Fe] $\sim$0 dex. 
Hill et al. (2011) discussed the possible contamination of the two populations by stars of the inner disk and halo and concluded that it is very little, although the situation is still uncertain (see also Bensby et al. 2011).
In a previous paper, Babusiaux et al. (2010)  found also kinematical differences among these two components: the metal poor component  shows a kinematics typical of an old spheroid (classical bulge), whereas that of the metal-rich component is consistent with a bar population (pseudo-bulge).
Bensby et al. (2011) measured detailed abundances of 26 microlensed dwarf and subgiant stars in the Galactic bulge, in particular the stars are all located between galactic latitudes $-2\,^{\circ}$ to $-5\,^{\circ}$, similar to Baade's window at (l,b)=($1\,^{\circ}$, $-4\,^{\circ}$). The analysis was based on high resolution spectra obtained with UVES/VLT. They also showed that the bulge MDF is double-peaked; one peak at [Fe/H]= -0.6$\pm 0.3$ dex, lower than the peak of the MP population of Hill et al. (2011), and one at [Fe/H]=+0.32$\pm 0.16$ dex. Clearly, the most recent observational evidence  points toward the existence of 
two main populations in the bulge, although the sample of dwarf stars needs to be substantially enlarged before drawing firm conclusions.

\section{The chemical evolution model}
Here we will try to model the two stellar populations (MP and MR), as described in the previous Section.  The chemical evolution model is similar to that adopted by
Cescutti \& Matteucci (2011), which is an upgraded version of that of Ballero et al. (2007), where a detailed description can be found. We remind here that the model can follow in detail the evolution of several chemical 
elements including H, D, He, Li, C, N, O, $\alpha$-elements, Fe and Fe-peak elements, s- and r-process elements.
The IMF is assumed to be constant in space and time and it 
is let to vary in order to test which one best fits the MDF. 
The star formation rate adopted for the bulge is a 
simple Schmidt law with exponent k=1. The efficienty of SF, namely the star formation rate per unit mass of gas, is let to vary from $\nu$=2 to 25 $Gyr^{-1}$.
In particular, to model the MP old spheroid component we adopt 
$\nu=25 Gyr^{-1}$, whereas for the MR pseudo-bulge component $\nu=2Gyr^{-1}$.
 This  model takes stellar feedback into account and compute the thermal energy injected into the interstellar medium (ISM) by SNe, as described in Ballero et al. (2007).

We assume that both stellar populations formed during episodes of  gas accretion: the law for gas accretion is assumed to be the same in both cases but the abundances of the infalling gas are different. We suppose that the gas which formed the MP component was primordial or slightly enriched from the halo formation, whereas the gas which formed the metal rich component was substantially enriched. In particular, the assumed chemical composition of the gas out of which formed the metal rich component has a [Fe/H]=-0.6 dex and all the abundance ratios reflect the composition of the gas forming the metal poor component at 
t$\sim$ 0.06 Gyr. It is worth noting that this particular chemical composition is similar to the composition of the gas in the innermost regions of the Galactic disk, near 2 kpc,  at an age of $\sim$2Gyr. The reason for the greater age
is that the inner disk must have evolved more slowly than the classical bulge, with a lower star formation rate. Therefore, the MR population could have started to form with a delay of 2Gyr relative to the MP one and out of gas of the inner disk.

The assumed gas accretion law is:
\begin{equation}
\dot{G}_{i}(t)_{inf}=A(r) X_{inf}\sigma(t_G)e^{-t/\tau}
\end{equation}
where $i$ represents a generic chemical element, $\tau$ is an appropriate collapse timescale fixed by reproducing the observed stellar metallicity distribution
function,  and $A(r)$ is a parameter
constrained by the requirement of reproducing the current total
surface mass density in the Galactic bulge ($\sigma_{bulge} \sim 1300 M_{\odot} yr^{-1}$), which in turn gives a total bulge mass of $\sim 2.0 \cdot 10^{10}M_{\odot}$ for a bulge radius of $R_B=2$kpc and surface mass density distribution following a Sersic profile (see Ballero et al. 2007). 
 Finally, $X_{inf}$ are the abundances of the infalling gas, considered constant 
in time and $t_G$ is the Galactic lifetime (13.7 Gyr). In particular, the abundances of the infalling gas are considered either primordial or slightly enriched at the level of the average metallicity of the halo stars ($<[Fe/H]> =-1.5$ dex). This second option is justified if we think that the bulge stars formed out  of gas lost from the halo and/or the inner thick-thin-disk.

The IMFs adopted here are: i) the one suggested by Ballero et al (2007):
\begin{equation}
\phi(m) \propto m^{-(1+x)}
\end{equation}
with $x=0.95$ for $M > 1 M_{\odot}$ and $x=0.33$ for $M < 1
M_{\odot}$ in the mass range $0.1-100M_{\odot}$.    
ii) The normal Salpeter IMF (x=1.35) in the mass range  $0.1-100M_{\odot}$, and
iii) the two-slope Scalo (1986) IMF, as adopted in Chiappini et al. (1997, 2001), with x=1.35  for $m< 2M_{\odot}$ and x=1.7 for $m \ge 2M_{\odot}$, always in the mass range  $0.1-100M_{\odot}$.

\subsubsection{Nucleosynthesis and stellar evolution prescriptions}
For the evolution of $^{7}$Li we have followed the prescriptions of Romano et al. (1999)  who predicted the evolution of Li abundance in the Galactic bulge.
The main Li producers assumed in that model are: i) core-collapse SNe, ii) massive-AGB stars, iii) C-stars, iv) novae and v) cosmic rays. We address the reader to this paper for details.

For all the other elements we have adopted the same yields as 
in Cescutti \& Matteucci (2011).
  In particular, detailed nucleosynthesis prescriptions are taken from: 
      Fran\c cois et al. (2004), who made use of widely adopted
      stellar yields and compared the results obtained by including
      these yields in a detailed chemical evolution model with the
      observational data, with the aim of constraining the stellar
      nucleosynthesis.  For low- and intermediate-mass ($0.8 -
      8M_{\odot}$) stars, which produce $^{12}$C and  $^{14}$N , 
      yields are taken from the standard model of van 
      den Hoek \& Groenewegen (1997) as a function of the initial
      stellar metallicity. Concerning massive stars ($M>10M_{\odot}$),
      in order to best fit the data in the solar neighbourhood, when
      adopting Woosley \& Weaver (1995) yields, Fran\c cois et
      al. (2004) found that Mg yields should be increased in
      stars with masses $11-20M_{\odot}$ and decreased in stars larger
      than $20M_{\odot}$, and that Si yields should be slightly
      increased in stars above $40M_{\odot}$. In the range of massive
      stars we have also adopted the yields of Maeder (1992) and
      Meynet \& Maeder (2002) containing mass loss.  The effect of
      mass loss is visible only for metallicities $\ge Z_{\odot}$.
      The use of these yields is particularly important for studying
      the evolution of O and C, the two most affected elements (see
      McWilliam et al. 2008; Cescutti et al. 2009). For  Ba,
      we use the nucleosynthesis prescriptions adopted by Cescutti et al. 
      (2006) to best 
       fit the observational data for this neutron capture element 
      in the solar vicinity; the same nucleosynthesis prescriptions give also
      good results when applied to dwarf spheroidals (Lanfranchi et al. 2006)
      and to the Galactic halo using an inhomogenous model (Cescutti 2008).
      In particular, we assume that the s-process fraction of 
      Ba is produced in low mass stars ($1-3M_{\odot}$, see Busso et al. 2001), 
      whereas the r-process 
     fraction of Ba originates from stars in the range $12-30M_{\odot}$.

\section{Results} 
\subsection{The two main populations}
We run several numerical models by varying the most important parameters: the IMF, the efficiency of star formation and the timescale
 and chemical composition of the infalling gas. We have found that the best models,  have the following characteristics: i) the 
MP population is obtained by means of a very efficient star formation ($\nu=25 Gyr^{-1}$) and a very short timescale for infall 
($\tau=0.1$ Gyr), as in our previous papers (e.g. Ballero et al. 2007 and Cescutti \& Matteucci, 2011) but with a Salpeter IMF, 
less flat than the IMF suggested in Ballero et al. (2007). The reason for this choice is due to the fact that with a Scalo (1986)
 IMF the peak of the MDF of the MP population occurs at [Fe/H]$\sim$-0.6 dex, too low compared with the observed one. On the other 
hand, the Ballero et al. (2007) IMF predicts a peak at a too high metallicity, [Fe/H]$\sim$ 0. The model with Salpeter IMF, instead,
 can well reproduce the MDF of the MP population. 
The chemical composition of the infalling gas is assumed to be slightly pre-enriched ($<[Fe/H]>$ $\sim -1.5$ dex). The infall of 
primordial gas, in fact, would predict too many metal poor stars in the MDF.
ii) The MR population is obtained with a less efficient star formation ($\nu=2Gyr^{-1}$) and a longer infall timescale ($\tau=3$ Gyr), 
the Salpeter IMF and enriched infall. In particular, for the infalling gas we assume the chemical composition corresponding to the gas 
at an age of 0.06 Gyr since the beginning of formation of the MP population. By the way, this composition represents also the metallicity
 of the gas in the very inner disk at an age of 2 Gyr, as predicted by chemical evolution models of the Milky Way disk (Cescutti \&  al. 2007). 
The predicted metallicity distribution of our best model for the metal poor population (MP in Table 1) is in reasonable agreement with the 
observed one (see Figure 1). In Figure 2  we show the same MDFs of Figure 1 but convolved with a gaussian taking into account an average 
observational error of 0.25 dex, in agreement with  Hill et al. (2011). In Figure 3 we show our predictions for the MDFs of the MP and MR 
populations as functions of [Mg/H], and also in this case we find a good agreement with data. In Figure 4 we show the same distributions as 
in Figure 3 but convolved with a gaussian taking into account an observational error of 0.20 dex.

In Figure 5 we show the predicted [Mg/Fe] for the two populations. Clearly, in the MP population the majority of stars has a high
 [Mg/Fe] roughly constant for a large range of [Fe/H]: in particular, the [Mg/Fe] ratio starts declining at [Fe/H]$\ge$ -0.3 dex. 
In the MR population, instead, the [Mg/Fe] varies from +0.2 dex to -0.1 dex in a [Fe/H] range of [-0.5 -- +0.7] dex. This lower [Mg/Fe] is 
due to the fact that this population formed out of pre-enriched gas where the pollution from Type Ia SNe was already present. We
 note that the predicted [Mg/Fe] is slightly high at low metallicities.This is probably due to the assumed Mg yields in massive stars.
 The Mg yields are, in fact, still quite uncertain.
By lowering the Mg yields the predicted curve would run lower than it is now but it would not change its shape.
We have computed the mean Fe abundance, $<[Fe/H]>$, for the two populations: for the MP one we find $<[Fe/H]>$=-0.26 dex in very good
 agreement with Hill \& al. (2011), whereas for the MR one we find 
$<[Fe/H]>$=+0.26 dex. This difference in the mean Fe abundance of the two populations could be interpreted as a gradient itself, 
as suggested by Babusiaux et al. (2010), although it is not clear how the stars of the two populations are spatially distributed 
and mixed.
It is interesting to note that we do not find any development of a galactic wind during the formation of both components, and this 
is due to the deep Galactic potential well in which the bulge is sitting, at variance with what is found for an elliptical galaxy 
of the same mass as the Galactic bulge, which sits in a shallower potential well (Pipino \& Matteucci, 2004).

\begin{figure}[htb]  
    \begin{center}  
 \includegraphics[width=0.48\textwidth]{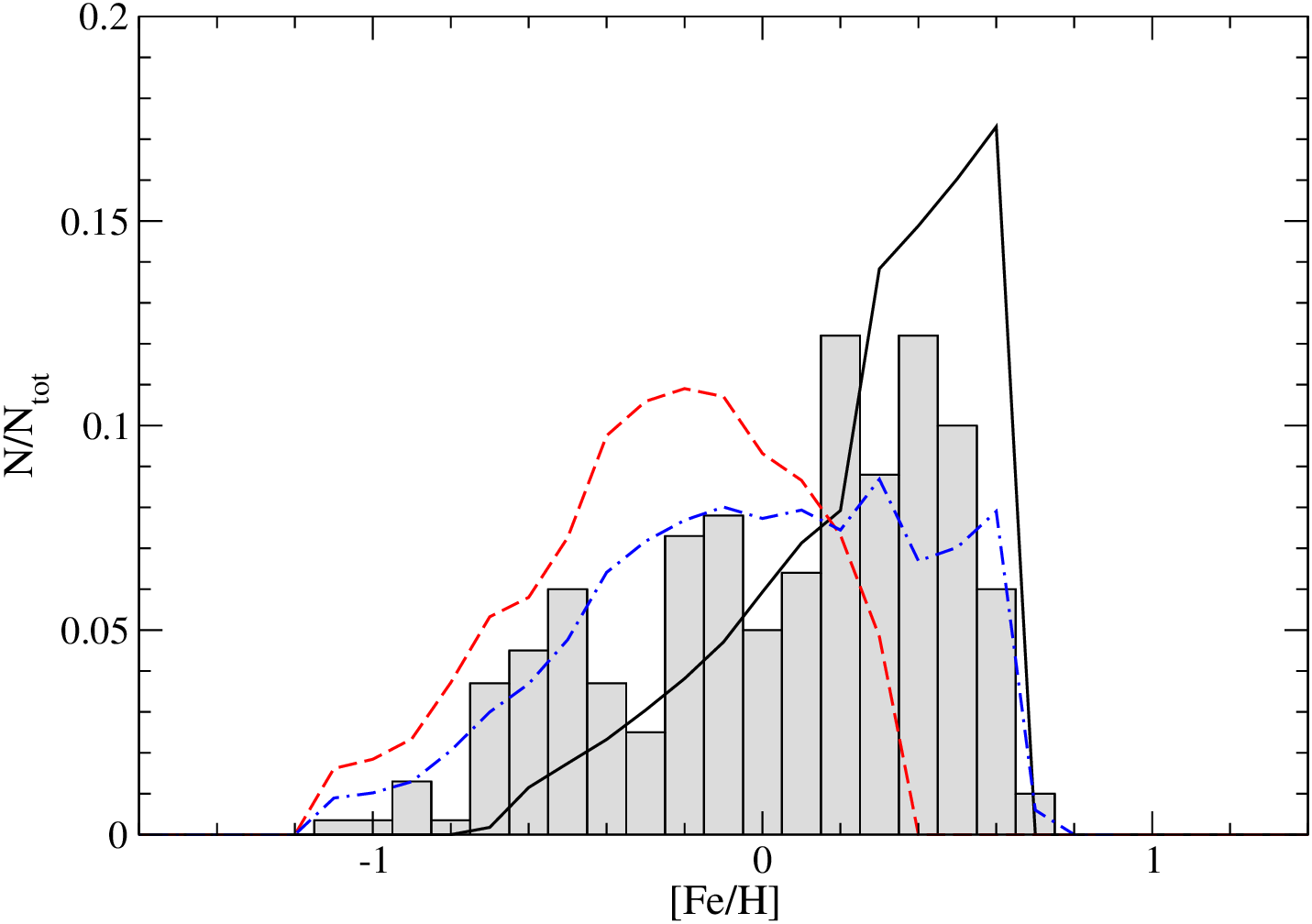}
    \caption{The predicted MDF for the two populations as functions of [Fe/H]: MP (dashed red line) and MR (continuous black line), compared to the data of Hill et al. (2011). The sum of the two distributions is also shown (blue dashed-dotted line).
     }
    \label{} 
    \end{center}
    \end{figure}

\begin{figure}[htb]  
    \begin{center}  
    \includegraphics[width=0.48\textwidth]{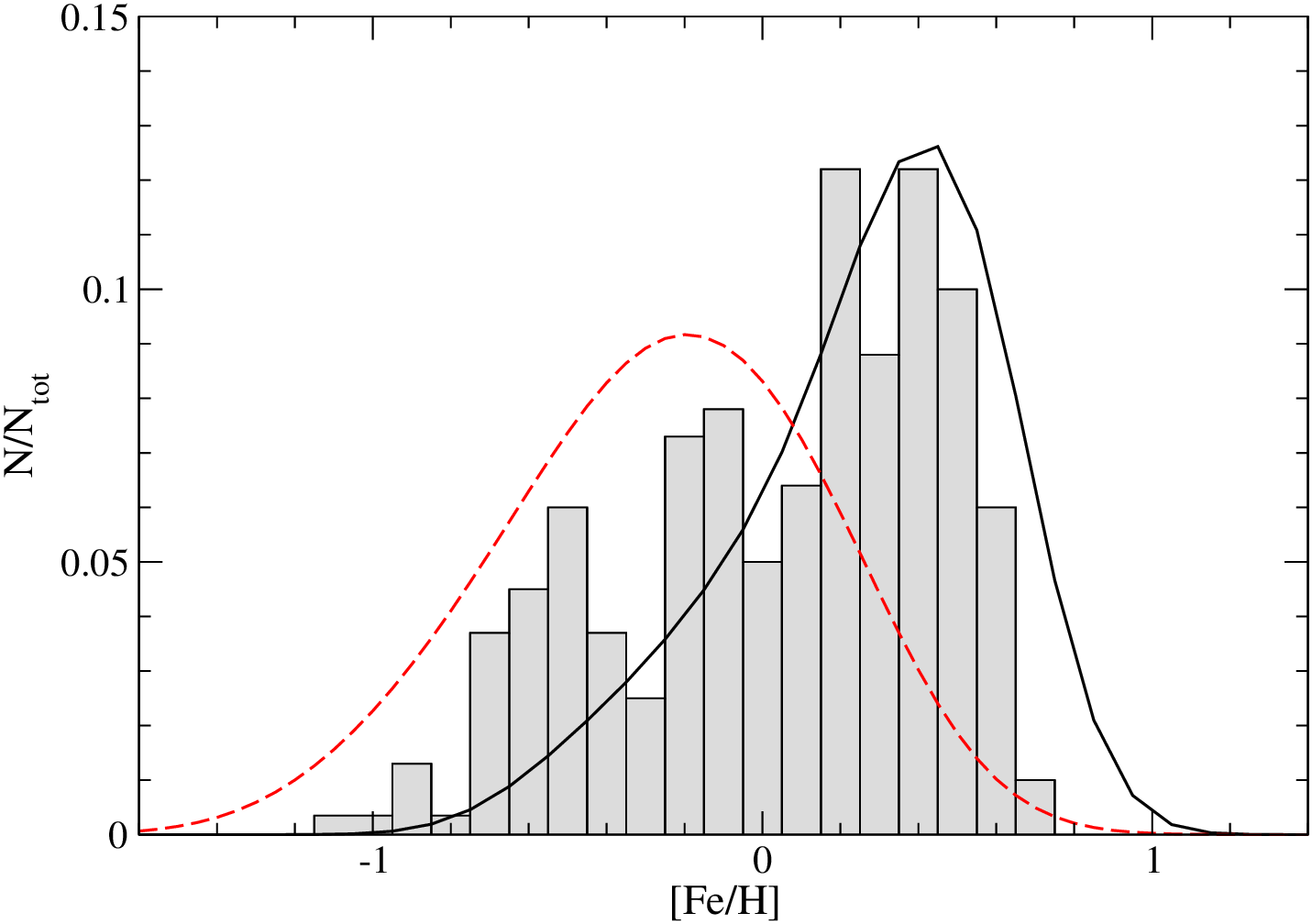}
    \caption{The predicted and observed MDF for the two populations as in Figure 1 but the MDFs have been convolved with a gaussian to take into account an observational error of 0.25 dex.
     }
    \label{} 
    \end{center}
    \end{figure}

\begin{figure}[htb]  
    \begin{center}  
\includegraphics[width=0.48\textwidth]{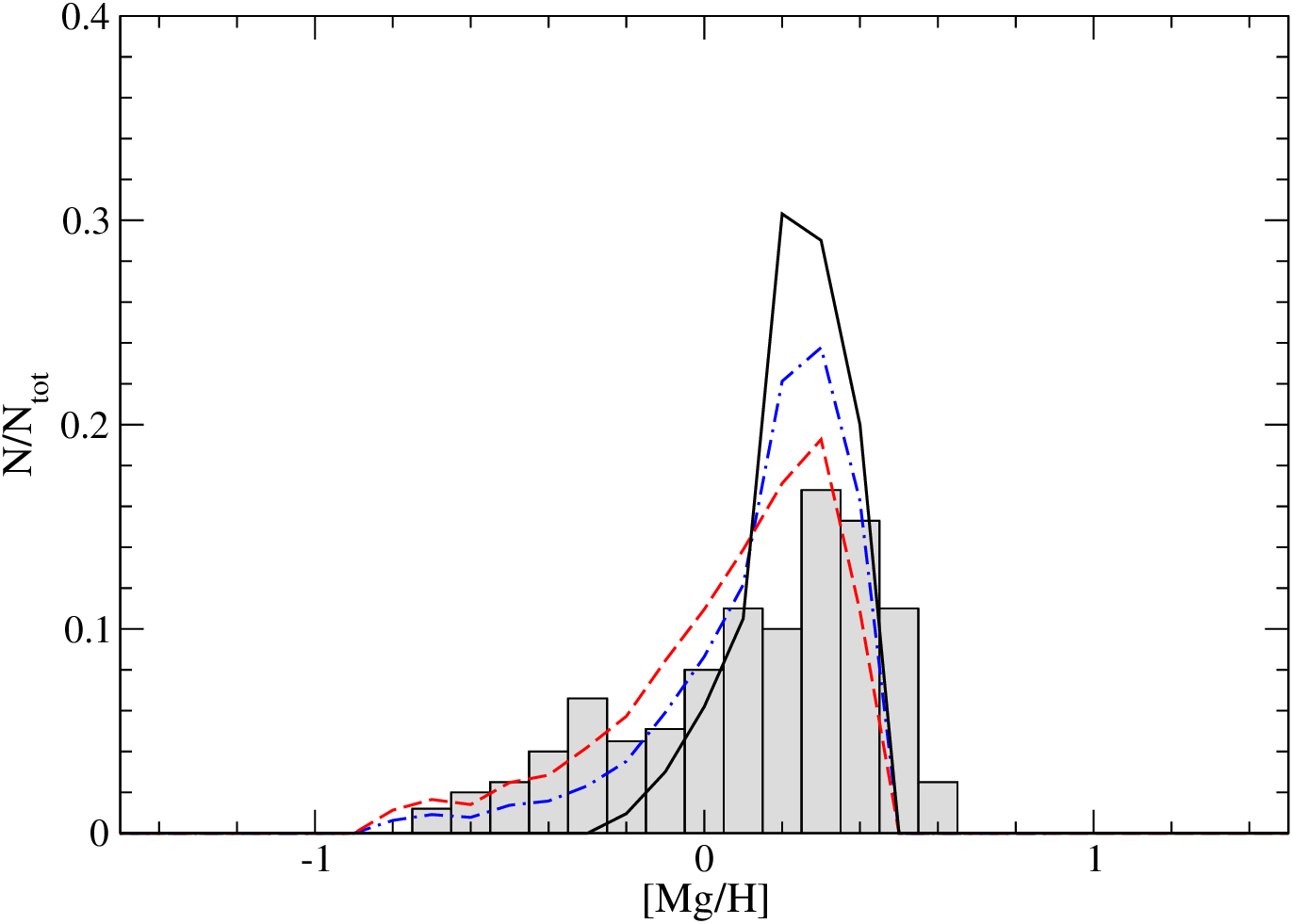}
    \caption{The predicted MDF for the two populations: MP(red dashed line) and MR (black continuous line) as functions of [Mg/H]. The data are from Hill et al. (2011). The sum of the two distributions is also shown (blue dashed-dotted line).  }

    \label{} 
    \end{center}
    \end{figure}

\begin{figure}[htb]  
    \begin{center}  
    \includegraphics[width=0.48\textwidth]{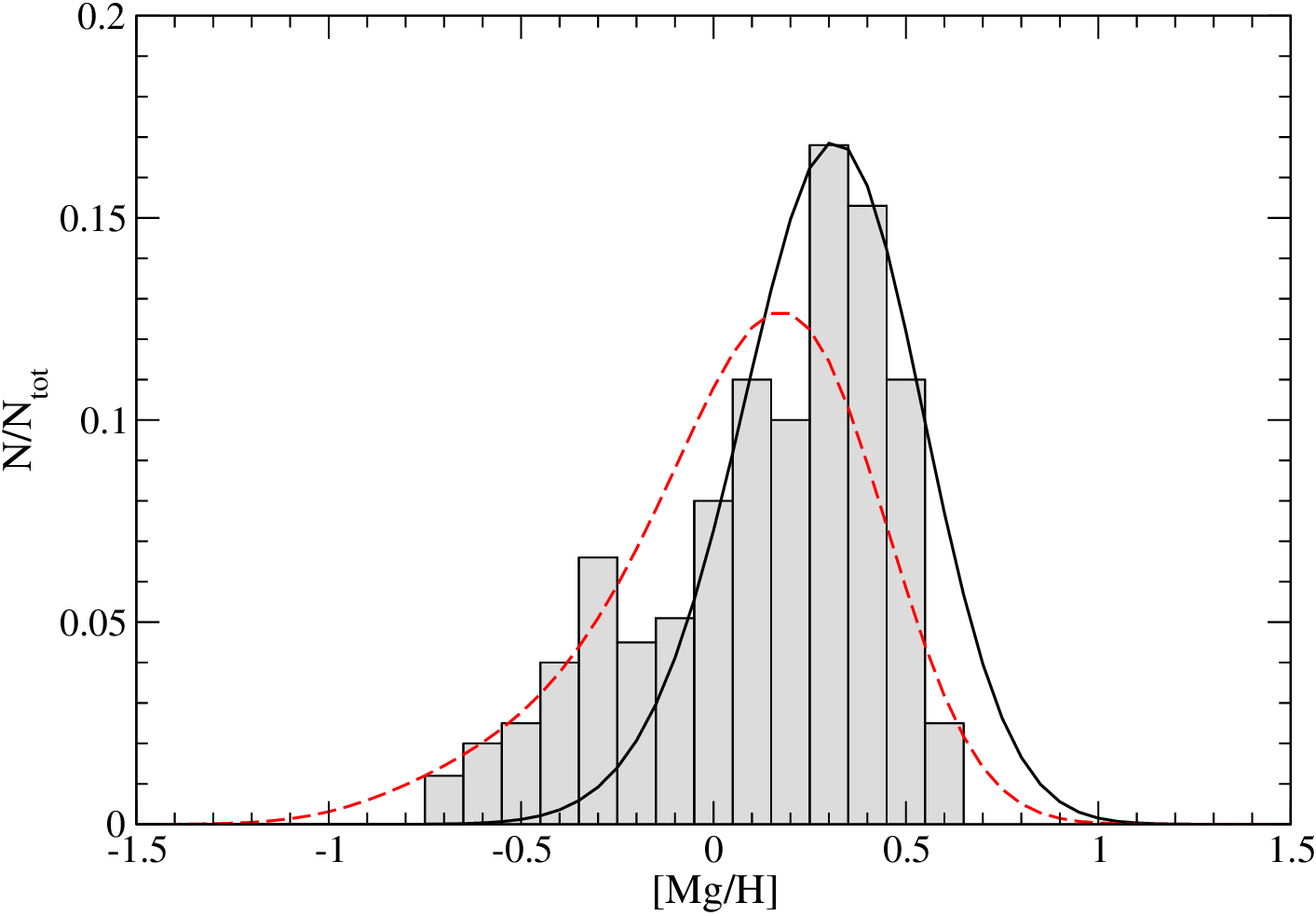}
    \caption{The predicted MDF for the two populations as a function of [Mg/H] as in Figure 3 but convolved with a gaussian corresponding to an observational error of 0.2 dex.
}
    \label{} 
    \end{center}
    \end{figure}

\begin{figure}[htb]  
    \begin{center}  
    \includegraphics[width=0.48\textwidth]{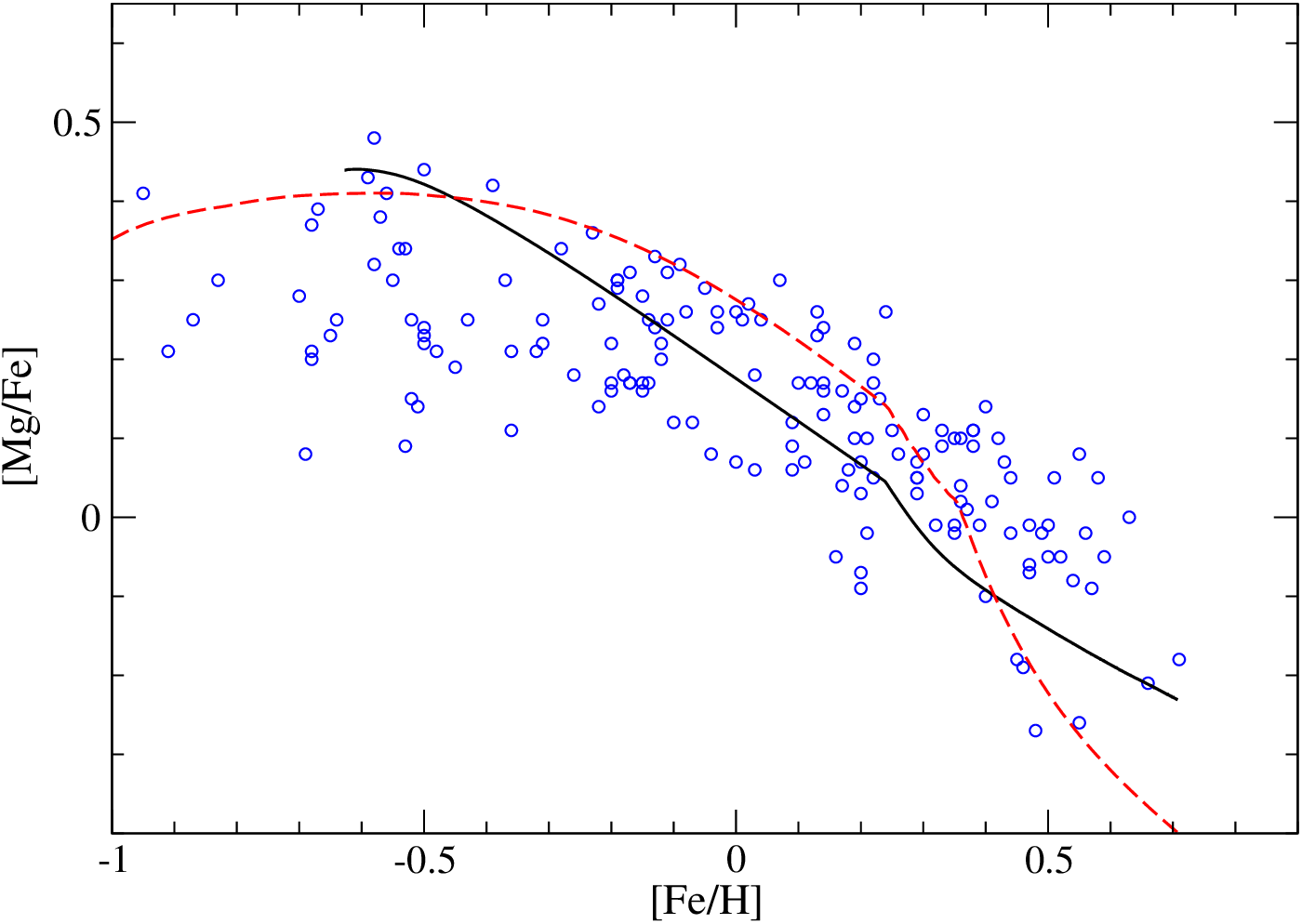}
    \caption{The predicted [Mg/Fe]vs. [Fe/H] for the MP and MR populations: MP (dashed red line) and MR (continuous black line), compared with data from Hill et al. (2011) containing both populations.}
    \label{} 
    \end{center}
    \end{figure}

\subsection{Possible abundance gradient in the metal poor population?}
The existence of abundance gradients in the Galactic bulge is a very important issue. Minniti et al. (1995) suggested the existence of an abundance gradient in the inner 
2 kpc but more recent analyses did not confirm this finding (Ramirez et al. 2000; Rich et al., 2007; Johnson  et al. 2011, 2012; Rich et al. 2012). In particular, 
these latter studies suggested the absence of a gradient from the Galactic center out to the Baade's window.
However, Zoccali et al. (2008;2009) found an abundance gradient along the bulge minor axis, as one moves from Baade's window to b=$-12\,^{\circ}$.  Such a gradient could
 be due to the formation of the bulge by dissipational collapse with the chemical enrichment being faster in the innermost regions. 
Here we test the idea that there could be a gradient inside the MP bulge population. We have computed then two models describing two different sub-populations of the metal
 poor component: a) the model for the innermost region has the same parameters as those adopted for the MP population ($\nu=25Gyr^{-1}$; $\tau$=0.1 Gyr; Salpeter IMF) but 
restricted to the inner 0.6 kpc, whereas the model for the outer region (from 0.6 to $\sim$ 2 kpc) has a lower star formation efficiency ($\nu=10Gyr^{-1}$), the same 
timescale for infall and the same IMF as the inner population. In Table 1 we summarize the model parameters for the four populations: 1) the MP old spheroid population, 2) 
the MR bar population, 3) the innermost sub-population of the MP (IMP) and 4) the outermost sub-population of the MP (EMP).
The predicted MDFs for the two sub-populations (IMP and EMP) are shown in 
Figure 6, where they are compared with the observed global MDF.
In Figure 7 we show the same MDFs but convolved with a gaussian with an average observational error of 0.25 dex, while in Figure 8 we present the resulting MDF obtained by 
summing the two MDFs of Figure 7. As one can see, the resulting MDF coincides practically with that of the MP population.
The two distributions are very similar although they peak at different [Fe/H] values. In particular, the difference between the mean Fe abundance of the IMP and that of the 
EMP is $\Delta<[Fe/H]>$=-0.145 dex.
Zoccali et al. (2008;2009) 
suggest  $\Delta<[Fe/H]>=-0.15$ dex going from b=$-4\,^{\circ}$ up to b=$-12\,^{\circ}$ (namely from 600 pc up to $\sim$ 1.6 kpc, in terms of galactocentric distance), in very
 good agreement with our prediction.
Our numerical models have shown that sub-populations with a larger 
gradient are not compatible with the observed MDF. In fact, a larger gradient would imply a larger difference between the predicted peaks of the MDFs of the two sub-populations, which is not observed. However, we cannot exclude the existence of a small gradient even between the Galactic center and the Baade's window.
\begin{table*}
\begin{center}
\begin{tabular}{|c|c|c|c|}
 \hline
  Model    &     $\nu$ ($Gyr^{-1}$) &    $\tau$(Gyr)   &  IMF \\  
 \hline
   MP     &    25   &   0.1     &  Salpeter  \\
 \hline
   MR     &    2    &   3.0     &  Salpeter     \\
\hline
   IMP       &   25   &   0.1  &  Salpeter    \\
\hline
   EMP       &  10   &   0.1   &  Salpeter    \\
\hline
\end{tabular}
\caption{Model parameters}
\end{center}
\label{tabmodels}
\end{table*}

\begin{figure}[htb]  
    \begin{center}  
    \includegraphics[width=0.48\textwidth]{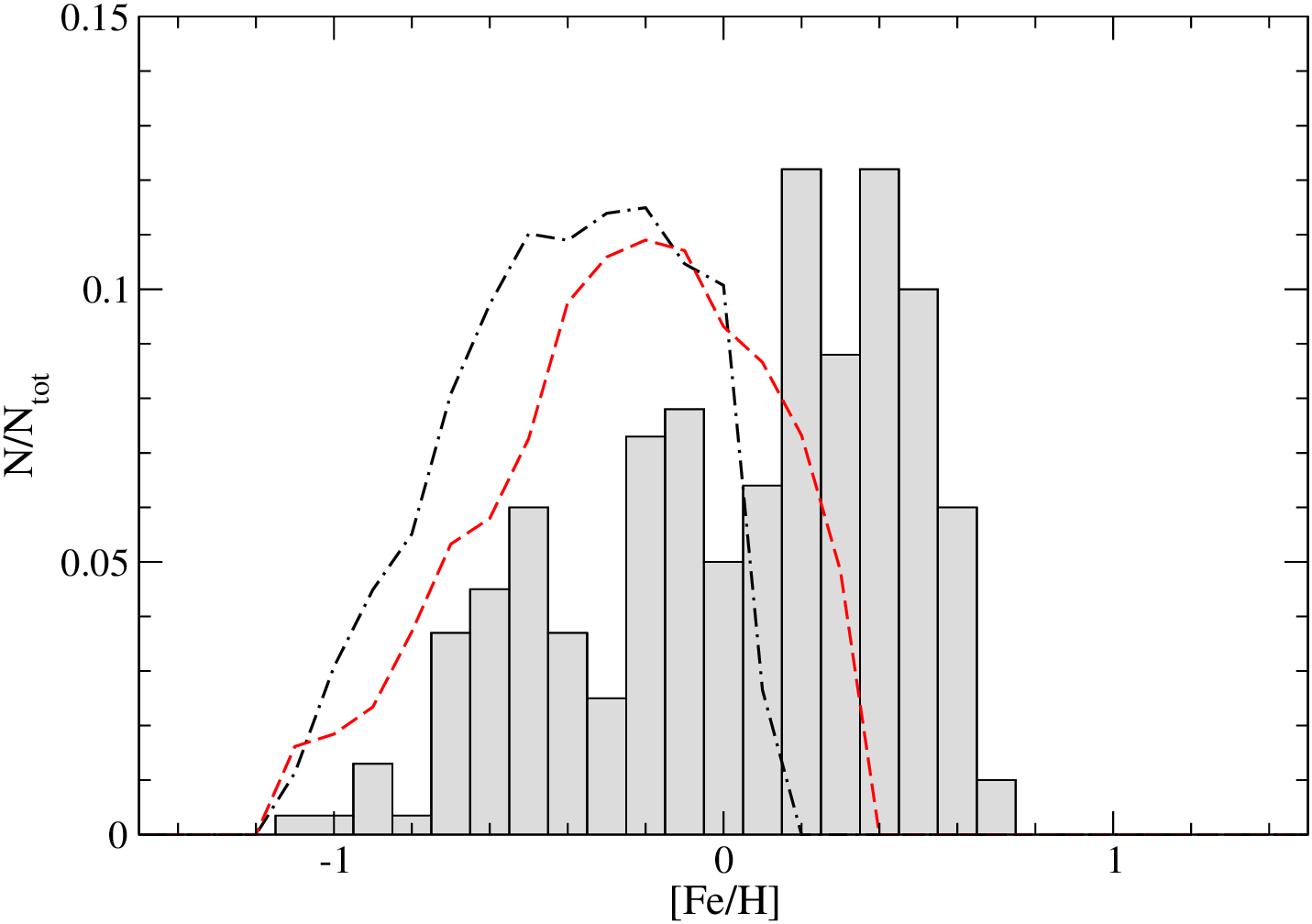}
    \caption{The predicted MDFs for the two possible sub-populations: IMP(red dashed line) and EMP (black dashed-dotted line) of the 
metal poor bulge population. The data are from Hill et al. (2011).
      }
    \label{} 
    \end{center}
    \end{figure}

\begin{figure}[htb]  
    \begin{center}  
    \includegraphics[width=0.48\textwidth]{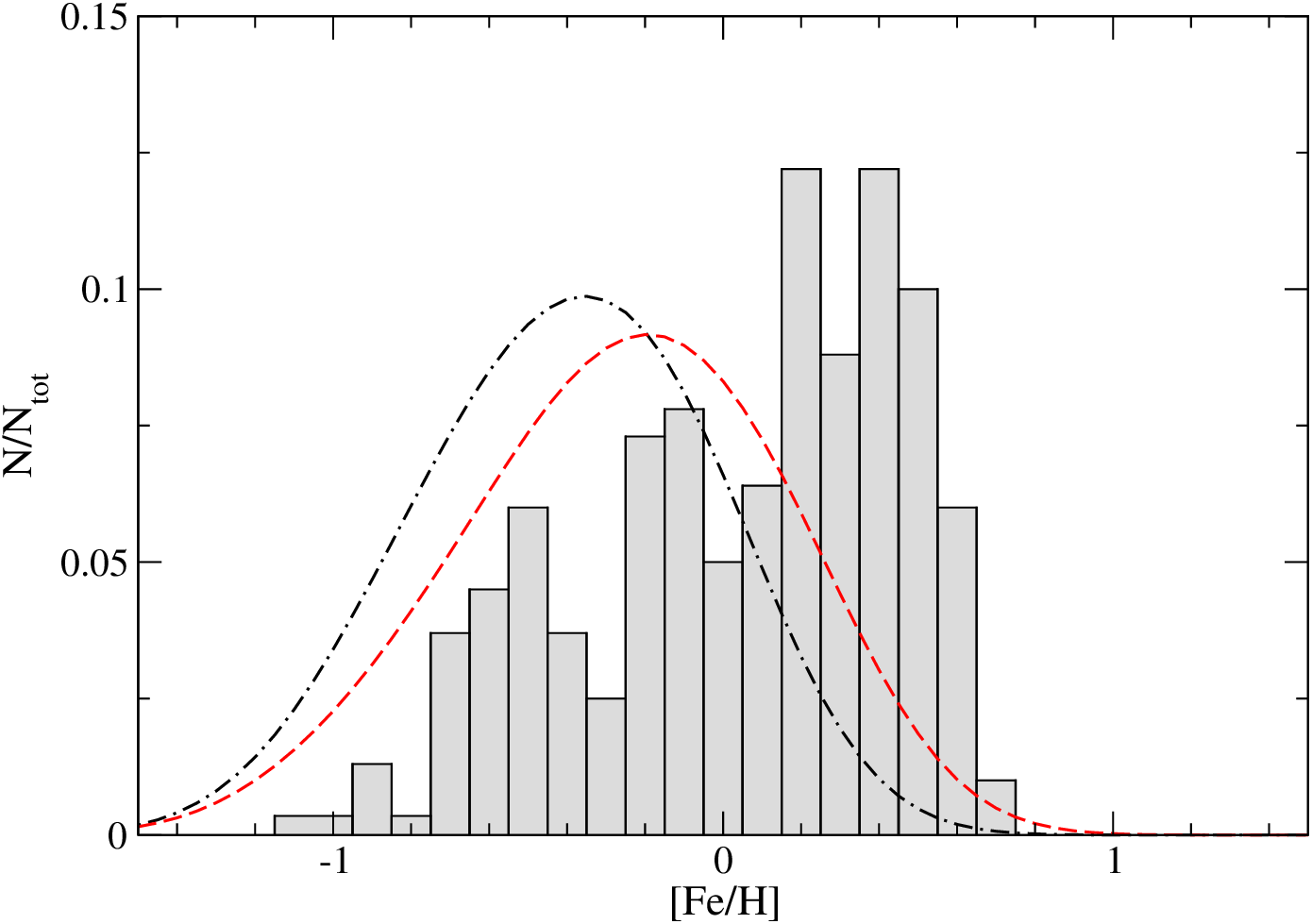}
    \caption{The predicted MDFs for the two possible sub-populations of the 
metal poor bulge population as in Figure 6, but convolved with a gaussian with an average observational error of 0.25 dex.}
    \label{} 
    \end{center}
    \end{figure}

\begin{figure}[htb]  
    \begin{center}  
    \includegraphics[width=0.48\textwidth]{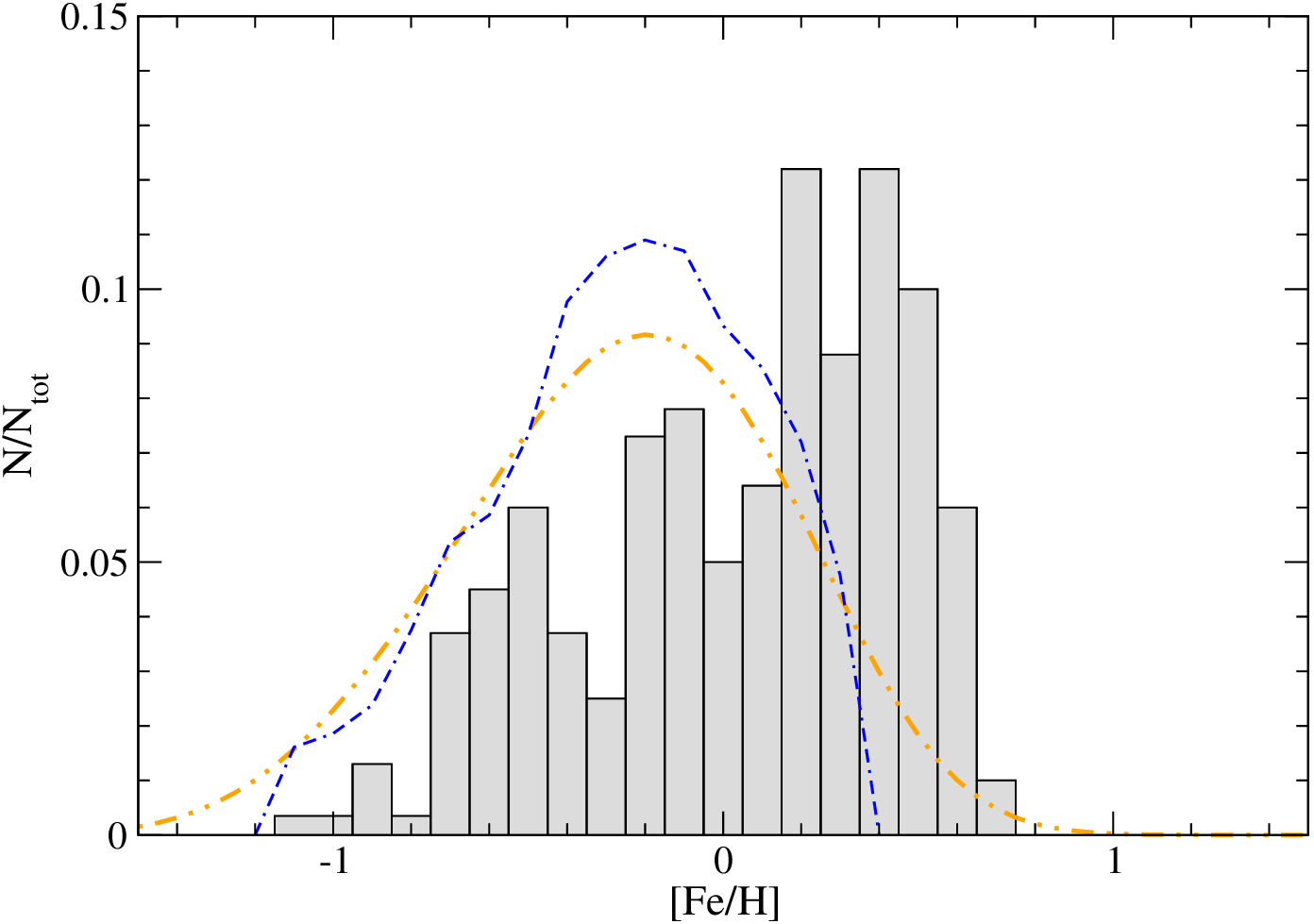}
    \caption{The predicted global MDF obtained by summing 
the EMP and IMP populations (dashed-dotted curve).
The global MDF is shown also in the gaussian convolved form (dashed-two-dotted curve).} 

    \label{} 
    \end{center}
    \end{figure}

We have computed the expected gradients for several chemical elements (Fe, Mg, O, Si, S and Ba) due to the differences between the average abundances of the two main 
bulge populations (MP-MR),  as well as those due to the differences in the average abundances in the sub-populations IMP and EMP. The results are shown in Table 2. 
For the gradients between MP and MR and IMP and EMP, we simply show the difference between the average abundances in their MDFs.

\begin{table*}
\begin{center}
\begin{tabular}{|c|c|c|c|c|c|c|}
 \hline
  Model    &    $\Delta <[Fe/H]>$  &    $\Delta <[Mg/H]>$   &  $\Delta <[O/H]>$ & $\Delta <[S/H]>$ &     $\Delta <[Si/H]>$ &  $\Delta <[Ba/H]>$  \\  
 \hline
   MP-MR      &   -0.521 dex   &   -0.232 dex     &  -0.214 dex &  -0.350 dex & -0.325 dex & -0.270 dex \\
 \hline
   EMP-IMP     &    -0.145 dex  &  -0.142 dex &   -0.137 dex & -0.140 dex & -0.145 dex  & -0.156 dex \\
\hline
   bulge3\, (${0-1kpc})$       &   -0.36 dex $kpc^{-1}$   &     &   -0.29 dex $kpc^{-1}$ & &  & \\
\hline
   bulge3\, $({0-0.5kpc})$       &   -0.26 dex $kpc^{-1}$   &      &   -0.21dex$kpc^{-1}$ & &   & \\
\hline
\end{tabular}
\end{center}
\caption{Differences among the mean abundances in the different stellar populations, as predicted by the 
different models (see text); they can be interpreted as gradients. Only for model {\it bulge3} we show the real gradient as $\Delta<[X/H]> / \Delta R$ and we compute
 it for two  distance ranges from the Galactic center  (0-1kpc and 0-0.5kpc).}
\label{tabmodels}
\end{table*}  
\subsection{The evolution of Li abundance in the gas of the bulge}

We have computed the evolution of the abundance of $^{7}$Li in the 
gas out of which the two main populations (MP and MR) formed. The reason for this is that recently Bensby et al. (2010; 2011) measured the Li abundance in 
several microlensed dwarfs and subgiant stars. Their data are plotted together with our model predictions in Figure 9. We are speaking of Li evolution in the
 gas and not in the stars because Li is easily destroyed inside stars and a galactic chemical evolution model aims at reproducing the upper envelope of the data
 in the plot logN(Li) vs. [Fe/H]. This procedure is commonly applied to the Li abundance data in the solar neighbourhood stars. The nucleosynthesis prescriptions
 adoptehere  correspond to those of {\it model C} of Romano et  al. (1999 and references therein). In this model, the contribution from core-collapse SNe to $^{7}$Li 
production is decreased by a factor of 2 relative to the predicted yields (Woosley \& Weaver 1995) and Li is mainly produced by massive-AGB stars and novae. As one can see, 
the predicted curve for the gas of the MP population very well reproduces the value of Li 
measured in the metal poor bulge dwarf  MOA-2010-BLG-285S which lies on the so-called Spite plateau observed in solar vicinity stars (Spite \& Spite, 1982). 
The other values for the Li abundance are all lower than that for  MOA-2010-BLG-285S but the stars are more metal rich and very likely the  $^{7}$Li in those objects 
has been depleted.
The initial value of  $^{7}$Li in Figure 9 has been assumed to be $logN_P$(Li)=2.2 as in Figure 5 of the paper of Bensby et al. (2010). This initial value corresponded, 
until a few years ago, to what we though was the primordial Li abundance. At the present time, the situation is more complicated since the primordial value for  $^{7}$Li,
 as estimated by WMAP (Hinshaw et al. 2009), is $logN_P(Li) \sim$ 2.6. No convincing explanation for this discrepancy has been found so far, and the most simple interpretation
 of this fact is that the primordial $^{7}$Li has been depleted in metal poor stars and for stars with [Fe/H] between -1.0 and -3.0 dex it must have been depleted by the same
 amount, thus  creating  the Spite plateau observed in the solar vicinity stars. For very low metallicities instead ($[Fe/H] < -3.0$ dex), the Li abundance could even further 
decrease (see Matteucci 2010 for a discussion and references therein).
\begin{figure}[htb]  
    \begin{center}  
    \includegraphics[width=0.48\textwidth]{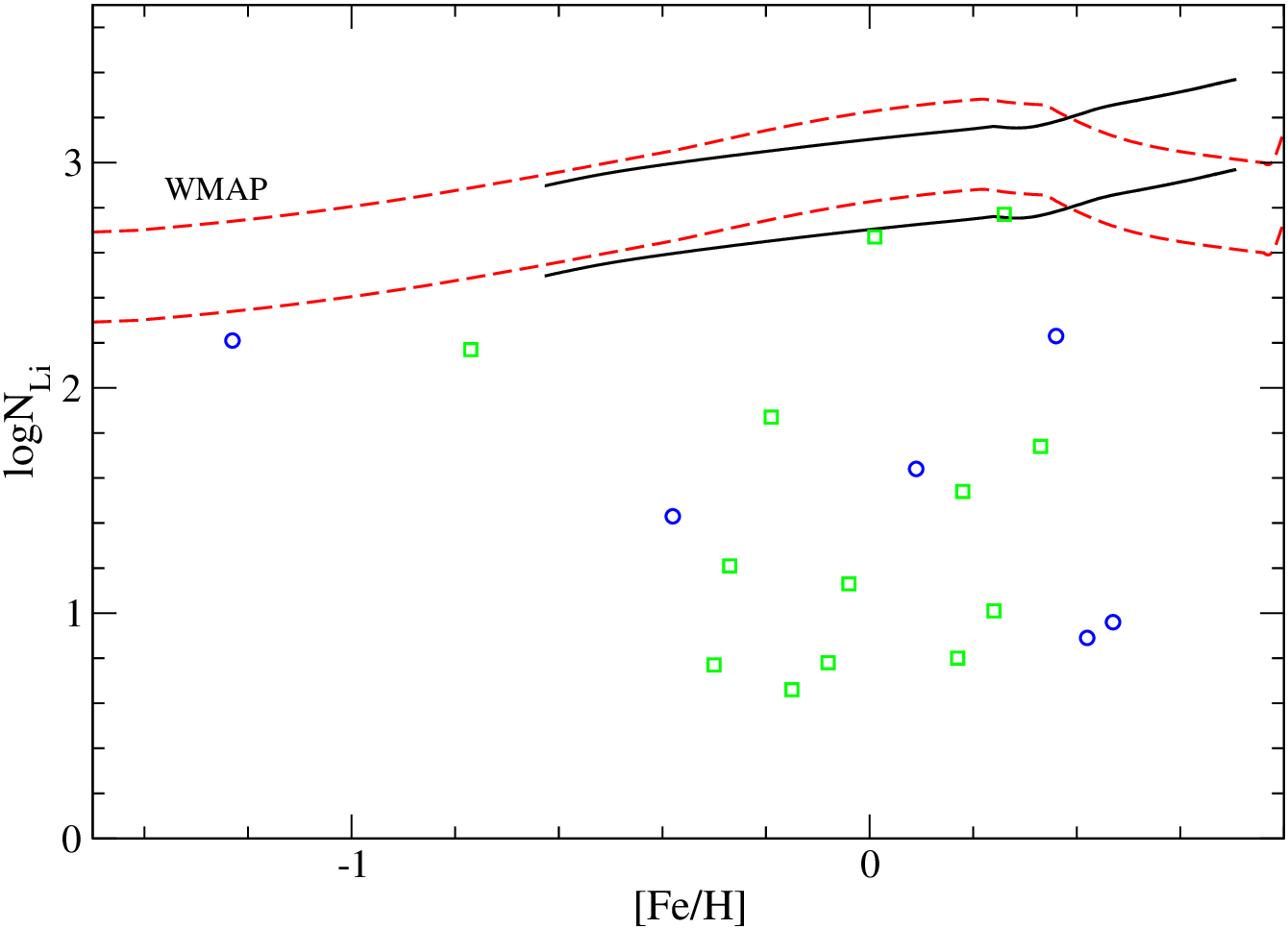}
    \caption{The predicted evolution of Li abundance for the gas in the MP and MR 
populations (see details in the text) compared with Li abundance determinations by 
Bensby et al. (2011), blue circle, and by Gonzalez et al. (2009), green square. 
The dashed red curve refers to the MP and the 
continuous black curve to the MR population. The star MOA-2010-BLG-285S corresponds 
to the point with [Fe/H]=-1.23 dex and logN(Li)=2.16.In the Figure we show also our predictions when the primordial Li abundance suggested by WMAP results is assumed.}
    \label{} 
    \end{center}
    \end{figure}

\section{A chemo-dynamical model for the bulge}
Here we summarize and extend the results obtained by Pipino et al. (2010), in particular those referring to the model labelled {\it bulge3}, in order to compare those
 results with the present ones. 
The model of Pipino et al. (2010) include gas hydrodynamics in 1D and follows the evolution of the abundances of O and Fe  and an extensive description can be found in 
Pipino et al. (2008;2010). In model {\it bulge3} a stellar mass of $2.28 \cdot 10^{10} M_{\odot}$ is formed on a time scale of $\sim$ 0.35 Gyr and with a Salpeter (1955) IMF,
 in excellent agreement with the results of Ballero et al. (2007) and Cescutti \& Matteucci (2011) and with the results of this paper concerning the formation of the MP population. 
It is worth noting that this chemo-dynamical model does not predict a bimodal population  but a continuous gradient inside the bulge. The bulge forms as a result of the collapse 
of primordial gas: in this kind of collapse at the beginning stars form everywhere, but as the collapse proceeds the gas accumulates towards the Galactic center where the metals 
tend to concentrate due to the gas, enriched by the very first stellar generations, falling towards the center. Under these conditions, if the collapse is not instantaneous an
 abundance gradient always form with the more metal rich stars sitting towards the center.  Due to the deep potential well of the Milky Way, no gas can escape from the bulge and 
the star formation stops just when the amount of gas is too low. This is in agreement with the results of this paper. The effective radius of the bulge is $\sim$ 1kpc, therefore
 by means of model {\it bulge3} we can compute the vertical gradients between the Galactic center and 1kpc; we find
a gradient of ${\Delta[Fe/H] \over \Delta R}$ = -0.36dex $kpc^{-1}$ and a gradient of ${\Delta [O/Fe] \over \Delta R}$= +0.07 dex $kpc^{-1}$. This positive gradient for the [O/Fe] 
ratio is due to the fact that star formation stopped earlier in outer region where the gas is rapidly lost because it falls towards the center. A shorter period of star formation
 means, in fact, higher [$\alpha$/Fe] ratios because of the less important contribution of the SNe Ia to the chemical enrichment. This predicted gradient in [Fe/H] is certainly
 measurable with high resolution spectroscopy of the bulge stars. Now, we can compare the gradient predicted by the chemo-dynamical model {\it bulge3} with the one obtained in
 this paper by assuming that the innermost region of the bulge evolved faster than the most external one. If we take the average 
$<[Fe/H]>$ of the two sub-populations and assume that the most metal poor one is
at $\sim$ 1.6 kpc from the center, then we find a gradient ${\Delta<[Fe/H]> \over \Delta R}\sim -0.1125 dex kpc^{-1}$. Altough less pronounced than the one derived  by means of the 
dynamical model, this gradient, if real, is still 
observable. The predicted gradients are reported in Table 2 where we show also the gradient predicted by {\it bulge3}  model from the center to 500 pc.

\section{Discussion and Conclusions}

Abundance ratios are useful tools to understand the timescale for the formation of different structures. The [$\alpha$/Fe] ratios measured so far in bulge stars have indicated that  a fraction of them formed on a short timescale, as indicated by the high and almost constant [$\alpha$/Fe] ratios for a large [Fe/H] range.
This means that only few stars belonging to this component formed out of gas polluted by Type Ia SNe, which occur with a time delay relative to core-collapse SNe (time-delay model).
Recently, an additional population of bulge stars with average [Mg/Fe]$\sim$0 and bar-like kinematics has been discovered, thus indicating that these stars must have formed either 
on a longer timescale than the other bulge stars or that they have formed out of gas already enriched and polluted by Type Ia SNe. Therefore, these indications seem to favor a 
complex scenario, with our bulge containing both the characteristics of a classical bulge and a pseudo-bulge. Abundance gradients have not been found in the innermost bulge region
 (up to b=$-4\,^{\circ}$), whereas from  b=$-4\,^{\circ}$ to  b=$-12\,^{\circ}$  Zoccali et al. (2008;2009) and Johnson \& al. (2011) found a gradient  along the bulge minor axis.
 In this paper, we propose a scenario which can explain 
the existence of the MP stellar population with possible abundance gradients inside it, together with the metal rich (MR) bar population.
Clearly, such a complex picture suggests to distinguish about two different types of gradients: the gradient formed inside the metal poor old bulge population by dissipational 
collapse,  and the gradient due to the differences in the average chemical abundances in the classical and pseudo-bulge populations. The gradient inside the classical bulge 
population should show a decrease in metallicity from the innermost to the outermost bulge regions.
On the other hand, the gradient between the classical bulge (MP) and the pseudo-bulge (MR) population should be more difficult to identify since the stars of the two populations 
could have been mixed at any Galactic latitude.

We have run different chemical evolution models to reproduce these different populations. Our conclusions can be summarized as follows:

\begin{itemize}
\item Both the MDF and the abundance ratios of the MP population can be reproduced by a classical chemical evolution model for the bulge; this model suggests a formation 
timescale of $\sim$ 0.1 - 0.3 Gyr, an IMF flatter than in the solar vicinity although less flat than previously suggested. In fact, the recent data can be well reproduced by
 a Salpeter (1955) (x=1.35) IMF. We assumed that the infalling gas forming this component was pre-enriched at the level of the average metallicity of halo stars ($<[Fe/H]>= -1.5$ dex).
 If this is true, we predict that stars with [Fe/H]$<$-1.5 dex should not exist in the Galactic bulge.

\item We also assumed that inside the classical bulge the most internal stars formed more rapidly than those
more external and to simulate this effect we simply assumed that the star formation efficiency was more rapid internally ($\nu=25Gyr^{-1}$) than externally ($\nu=10Gyr^{-1}$), 
thus mimicking a dissipative gravitational collapse. Then we computed the average abundances of these two distinct sub-populations and found that the sub-population higher 
on the galactic plane should be less metal rich, in agreement with what found by Zoccali et al. (2008;2009). We also run a chemo-dynamical model following a dissipational collapse 
for the bulge formation and predicted a gradient of ${\Delta [Fe/H] \over \Delta R}=-0.26 dex kpc^{-1}$ in the Galactocentric distance range 0-500pc. 
Babusiaux et al. (2010) suggested that the change in the metallicity distribution function  and in the kinematics as a function of metallicity  with increasing galactic latitude is 
due to the MR population disappearing while moving away from the plane.
If this is true any (residual) observed gradient at  b $< -6\, ^{\circ}$ must be an intrinsic property of the MP population. Here we showed
that such a gradient is expected in the monolithic dissipational assembly of the MP population.

\item Then, we run a model to explain the MR population: we assumed that these stars formed with a delay relative to those of the MP population and out of a substantially
 enriched gas and on a longer timescale of $\sim 3$ Gyr.
We compared our results with the MDF and the [$\alpha$/Fe] ratios of the MR population of Hill et al. (2011) and found a good agreement.

\item We predicted abundance gradients for Fe, Mg and also for O, S, 
Si and Ba between the MP and MR populations. The gradients (i.e. differences in the mean abundances of the two populations) we found are quite substantial (up to -0.7 dex for 
[Fe/H]) and certainly observable with high resolution spectroscopy.   
\item Finally, we presented predictions for the evolution of the abundance of $^{7}$Li in the gas out of which the MP and MR populations formed. We found a good agreement with 
the abundance of Li measured in a metal poor star MOA-2010-BLG-285S and suggested that to obtain this agreement one needs to decrease the  $^{7}$Li production by core-collapse SNe.
 A negligible contribution to Li by core-collapse SNe has also been recently suggested by Prantzos (2012).
\end{itemize}

Therefore, although the existence of a bimodal population is not yet proven without doubt, in agreement with Babusiaux et al. (2010) we suggest that from the chemical point of
 view it is possible the co-existence, in the Galactic bulge, of two main different stellar populations, one probably related to a fast gravitational collapse and the other to
 the existence of the bar, which appears to be a predominant feature in the bulge. Last but least, we have shown that abundance gradients inside the bulge population formed by
 gravitational collapse can also exist, thus suggesting the existence of even more than two stellar populations. However, only future high resolution observations of bulge stars 
will help to clarify this complex scenario. If no gradients will be found in the inner bulge then our conclusion will be that the stars formed  very fast
 (timescale $\le$ 0.1-0.3 Gyr) with no energy dissipation  and with the same high efficiency of star formation ($20-25 Gyr^{-1}$).

\begin{acknowledgements}
We are indebted to C. Flynn and M. Nonino for providing programs for handling the comparison between predictions and data.
We thank D. Romano, O. Gonzalez, R.M. Rich and M. Schultheis for their careful reading of the manuscript and very useful suggestions and M. Zoccali for many illuminating discussions. We also thank an anonymous referee for his/her suggestions which improved the paper.
\end{acknowledgements}

\end{document}